\documentclass[conference]{IEEEtran}
\IEEEoverridecommandlockouts
\usepackage{cite}
\usepackage{amsmath,amssymb,amsfonts}
\usepackage{algorithmic}
\usepackage{graphicx}
\usepackage{textcomp}
\usepackage{xcolor}
\usepackage{listings}
\def\BibTeX{{\rm B\kern-.05em{\sc i\kern-.025em b}\kern-.08em
    T\kern-.1667em\lower.7ex\hbox{E}\kern-.125emX}}

\newtheorem{definition}{Definition}
    
\begin{document}

\title{Emergent Software Service Platform and its Application in a Smart Mobility Setting\\
}

\author{
\IEEEauthorblockN{Christoph Knieke, Eric Nyakam,\\Andreas Rausch, Christian Schindler}
\IEEEauthorblockA{\textit{Technische Universität Clausthal} \\
\textit{Institute for Software and Systems Engineering}\\
Clausthal-Zellerfeld, Germany \\
Email: \{christoph.knieke, eric.douglas.nyakam.chiadjeu, \\ andreas.rausch, christian.schindler\}@tu-clausthal.de}
\and
\IEEEauthorblockN{Christian Bartelt, Nils Wilken}
\IEEEauthorblockA{\textit{Institute for Enterprise Systems} \\
\textit{University of Mannheim}\\
Mannheim, Germany \\
Email: \{bartelt, wilken\}\\@es.uni-mannheim.de}
\and
\IEEEauthorblockN{Nikolaus Ziebura}
\IEEEauthorblockA{\textit{Anaqor AG} \\
Berlin, Germany \\
Email:\\nikolaus.ziebura@anaqor.io}
}

\maketitle

\begin{abstract}
The development dynamics of digital innovations for industry, business, and society are producing complex system conglomerates that can no longer be designed centrally and hierarchically in classic development processes.
Instead, systems are evolving in DevOps processes in which heterogeneous actors act together on an open platform.
Influencing and controlling such dynamically and autonomously changing system landscapes is currently a major challenge and a fundamental interest of service users and providers, as well as operators of the platform infrastructures.
In this paper, we propose an architecture for such an emergent software service platform.
A software platform that implements this architecture with the underlying engineering methodology is demonstrated by a smart parking lot scenario.
\end{abstract}


\section{Introduction} \label{sec:introduction}

The runtime environment of current software systems, such as modern embedded systems or information systems, consists of a large number of complex software components that run on different hardware components in a distributed architecture.
In a distributed system architecture, each of these components can use and provide different applications or services over a network connection.
This environment is also often referred to as the \textit{Internet Of Things (IoT)} \cite{al2015internet}.
One important feature of an IoT runtime environment is that the connected components can be from very different domains (e.g., social events, transportation, home automation, etc.).

From a technical point of view, in such an environment, a software platform, which enables software service providers to offer their services and enables customers to make use of these services, is required.
We refer to such a software platform as a \textit{Platform Ecosystem}.
In this work, we focus on \textit{software services} as these software components.
A major challenge in the context of such Platform Ecosystems is the composition of the available software services so that they function together as a more complex, higher-value software component.

Due to the dynamic nature of an IoT environment, another major challenge for such Platform Ecosystems is to maintain their functionality also if runtime conditions suddenly change, which is often the case in IoT environments \cite{broring2016categorization}.
Recent self-adaptive software systems, which are able to adapt their architecture and behaviour automatically to changes in the environment to a certain extent, try to solve this challenge.
Such self-adaptive systems can be designed from existing software services, as proposed in the DAiSI component model \cite{rehfeldt2017component}.
However, such self-adaptive systems still have to be designed manually to a large extent.
Hence, we propose the concept of an \textit{Emergent Software Service Platform}, which is able to design software services from the set of available software services completely automatically at runtime.

This vision includes, that such an Emergent Software Service Platform has to have at least the following capabilities:
\begin{enumerate}
    \item It has to be able to elicit the current user requirements automatically at runtime.
    \item It has to be able to automatically compose a software service, which meets the elicited user requirements, from the set of available software services at runtime.
    \item It has to be able to execute an automatically composed software service at runtime and provide the result to the user of the Emergent Software Service Platform.
\end{enumerate}

The paper is structured as follows: 
Section~\ref{sec:relatedWork} gives a short overview on related work. 
In Section~\ref{sec:architecture}, we introduce our platform architecture for emergent software service composition. 
We apply our software platform on a mobility use case, described in Section~\ref{sec:appication}.
Finally, Section~\ref{sec:conclusion} concludes.

  

\section{Related Work} \label{sec:relatedWork}

Influencing and controlling dynamically and autonomously changing system landscapes is currently a major challenge and a fundamental interest of service users and providers, as well as of platform infrastructure operators.
For this reason, numerous platform and middleware technologies supporting controlled self-adaptation of dynamically adaptive systems have emerged in research and development over the last decade \cite{klus2014daisidynamic,Cardellini2017,hallsteinsen2012development,ANDRADE2021106505,burger2020elastic}. 
PORSCE II \cite{hatzi2013porsce} is one of the first semantic composition systems for web services with the particularity that it takes advantage of semantic information to improve the planning, as well as the composition of software components.
iServe \cite{villalba2015servioticy}, on the other hand, is not directly concerned with the composition of Web Services but describes a new and open platform for publishing web services to better support their discovery and use. 

However, although these system approaches already support both dynamic networking at runtime, self-adaptation, and openness, their reliable operation depends on the presence of a central logical entity (e.g., a platform operator and/or a standardization body).
Such an actor is currently needed to standardize data, services, and processes in such a way that component providers can specify semantically compatible interfaces on their basis.
Furthermore, strategies for functional and non-functional assurance of self-adaptive IT systems also require a central body to guarantee reliable cooperation among components.
Current technology platforms centrally specify the configuration rules according to which system components can network with each other.
However, the emergence of far-reaching emergent systems is severely restricted by such centrally anchored coordination mechanisms.




\section{Emergent Software Service Platform} \label{sec:architecture}

In this section, we describe a revised version of the architecture of a software service platform, which was introduced in \cite{wilken2020dynamic}, that has all the capabilities that were described above.

\subsection{Definitions}
First, we begin with defining the core concepts of an emergent software service platform in the context of an IoT runtime environment.

\begin{definition}[Software Service Description]
A software service description defines the required and provided interfaces of a software service instance.
\end{definition}

\begin{definition}[Software Service Instance]
A software service instance is a software entity that implements the interfaces that are defined by the corresponding software service description.
In addition, this software service entity is already deployed and ready for use.
\end{definition}

\begin{definition}[Process]
A process is a composition of software service descriptions to describe the composed system behavior, which can be executed by an execution engine.
\end{definition}

\begin{definition}[Software Service Platform]
A software service platform is a software platform that provides a library of software services (not software service instances) to potential users.
Providers that host software service instances can register their service instances on the platform with their corresponding software service descriptions.
\end{definition}

\begin{definition}[Emergence]
A software service platform is called emergent if it automatically and dynamically composes available software services to an executable software service in response to a trigger event (user requirement).
The executable software service is not predefined at design time and cannot be anticipated by the individual components.
\end{definition}

\subsection{Architecture}

\begin{figure*}
\centering
   \includegraphics[width=0.8\textwidth, angle=0]{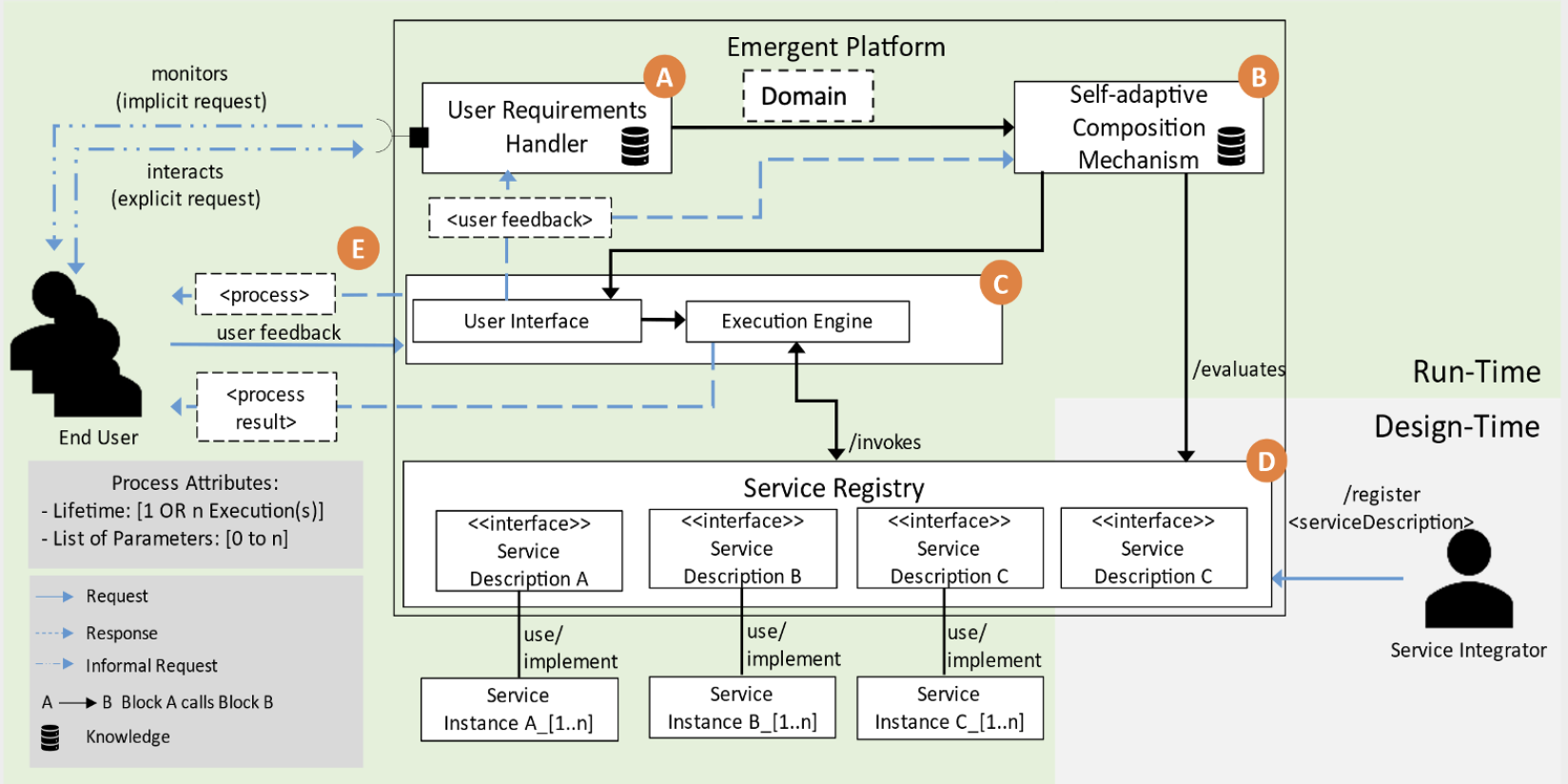}
   \caption{Platform Architecture.}
   \label{fig:Architecture}
\end{figure*}

The core idea of a dynamic adaptive IoT ecosystem is the reuse of software components and the ability to fulfill the expected system behavior by using emergent sequences of available software components to provide higher-value services.
Compared to the previous work \cite{wilken2020dynamic}, as shown in Figure \ref{fig:Architecture}, the architecture additionally includes the \textit{Domain} as an internal component of the platform. 
The emergent platform as a whole interacts with users to determine formal user requirements through interactions and monitoring (A).
The formal user requirement is then passed on to the composition mechanism (B) to compose a sequence of Software Service Descriptions to fulfill the expected system behavior demanded by the formal user requirement.
The Service Descriptions that are allowed to be used in the composition are registered in the Service Registry (D).
The composed sequence of components is forwarded to the Execution Engine (C).
The responsibility of this component is to call Service Instances that ``use/implement'' the given software components of the composed sequence, to incorporate necessary user feedback into the execution, and to return process results to the user.
The Domain is a central part of the architecture, as it is the foundational vocabulary to express user requirements and the foundation for a semantic description of the software components. 

\paragraph*{Requirements Handler}

The requirements handler component is responsible for the automatic elicitation and formalization of the current user requirements.
This can be done following two different modes of interaction with the user.
First, the user can explicitly state his/her user requirements in a request to the platform.
However, the majority of users are not capable of expressing their requirements in a formalized format.
The task of the requirements handler component is to convert the unformalized user requirements into a formalized format.
Second, there is the option of implicit user requirements recognition.
This means that the user does not proactively interact with the platform but the user requirements are extracted from a sequence of sensor measurements that the emergent software platform is able to record from the environment of the user.
In this case, the task of the user requirements handler component is to extract the current user requirements from the sequence of sensor measurements.

\paragraph*{Self-adaptive Composition Mechanism}

 The composition component handles the fulfillment of a user request as a planning problem.
 The user requirement is the goal of the planning problem.
 A sequence of software components is to be determined (the plan) to fulfill the expected behavior.
 As the goal and the available software components are semantically described in terms of concepts defined in the Domain a matching of (parts of the) user requirement to software components is possible without the need to explicitly describe compatibility of software components.

\paragraph*{Execution Engine}

The execution engine provides a browser-based flow editor to integrate external services into the platform via the concept of an interpretable flow.
In addition, it provides a runtime environment with an interpreter to execute those flows.
Each flow consists of a set of nodes, wired together to define the order of execution and the way data is transferred from node to node.
With the concept of service-nodes, flows can be combined into a new, more complex, (meta-) flow.
This is the base concept for the automatic creation of a flow based on a composition result.
For each service description in the service repository, there has to exist at least one flow tagged with the related action-reference.
Now, the composition result can be mapped to an executable meta-flow by finding a matching flow for each action and wiring them together in the order given by the composition result.

\paragraph*{Service Registry}

The Service Registry contains the Service Descriptions to be offered by the platform. This is primarily a set of unique service descriptions that can be associated with different Service Instances. The Service Registry manages both the location and the style of interaction with the Service Instances. For example, available Service Descriptions can be queried by the Self-adaptive Composition Mechanism during composition. In addition, new Service Descriptions can be added to the Service Registry or the associated instances can be invoked by the execution engine.

\section{Application to a Smart Mobility Setting} \label{sec:appication}

We will demonstrate the introduced architecture in the application domain of a smart mobility IoT ecosystem, other domains are also feasible, as the introduced architecture is not limited to the demonstration domain.
It consists of different services offered in the context of a parking lot that are able to be used on their own and give additional value when used in combination.
We have defined web services that can have an impact in a real-world parking lot and trigger physical actions.
In addition, we have defined services to place a reservation in a parking lot, charging of an electric vehicle, book a car wash, get tire pressure measurement, and get navigation directions to a specific spot in the parking lot.
The services are described as OpenAPI-specified REST-endpoints and prototypical implementations are available.
Based on the OpenAPI-specifications we have integrated the services as Service Descriptions in the Service Registry of the prototype.

\begin{figure}[b]
\centering
   \includegraphics[width=0.48\textwidth, angle=0]{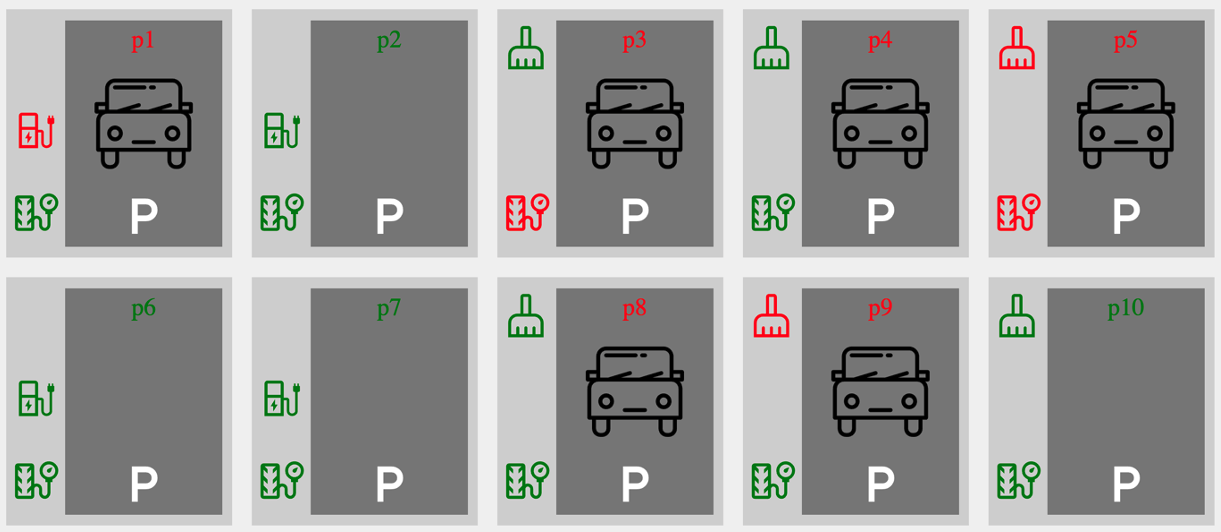}
   \caption{Parking Lot UI.}
   \label{fig:parkinglot}
\end{figure}

\begin{figure}[!ht]
   \centering
   \includegraphics[width=0.25\textwidth, angle=0]{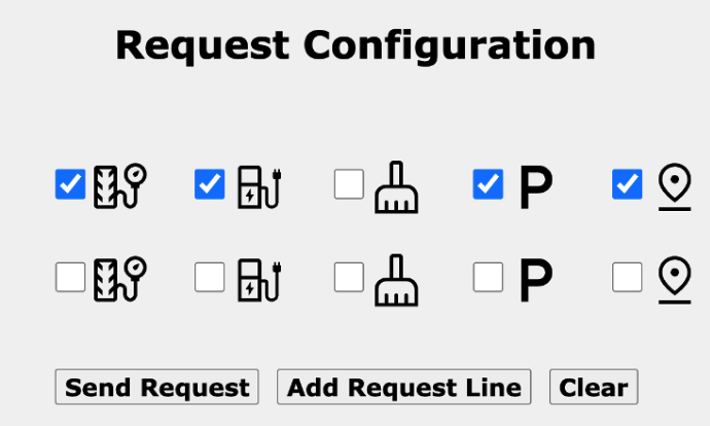}
   \caption{User Request UI}
   \label{fig:userreq}
\end{figure}

In addition to implementing all the components shown in the platform architecture, we have also implemented software components, a formal domain, the semantic integration of the software components in the service registry, a GUI illustrating a parking lot in Figure \ref{fig:parkinglot}, and an additional GUI part to accept explicit user requests, see Figure \ref{fig:userreq}.

For each parking spot in Figure \ref{fig:parkinglot}, icons indicate the available services (green) and the booked services (red). 
Figure \ref{fig:userreq} shows the Request Configurator, which an end user can interact with to express requests the platform should fulfill. Each icon on the UI represents a certain desire and each row refers to a certain request in the same parking lot (e.g., the selected  checkboxes in the first row mean, that tire pressure measurement and charging station are desired, the parking spot should be booked and the navigation directions to the same parking spot should be given).
The request will be formulated in a formal format, given in Figure \ref{fig:listing1}.
\begin{figure}[b]
\begin{lstlisting}[basicstyle=\footnotesize]
{"environment": [ 
 {"value":"", "type":"parkingid", "name":"p1"}, 
 {"value":"", "type":"operatorid", "name":"b1"},
 {"value":"", "type":"reservationnr", "name":"r1"},
 {"value":"", "type":"maxparkingtime", "name":"m1"},
 {"value":"", "type":"bookedservice", "name":"g1"}], 
"init": [], 
"goal": "(and (tirepressurecheck r1)
(bookeparking p1 r1 m1)
(navigation p1))"}
\end{lstlisting}
\caption{Formalized User Request.}
\label{fig:listing1}
\end{figure}
The request contains environment information, such as specified initial values for objects referenced in the request, an initial state, and the goal state.
Based on this request, the Self-adaptive Composition Mechanism is able to compute a Service Description sequence, given in Figure \ref{fig:listing2}.
The message of the composition forwards the objects and their initial values and the service description sequence toward the executing part of the platform, which selects suitable Service Instances for the descriptions and calls them according to the composition plan.

\begin{figure}[t]
\begin{lstlisting}[basicstyle=\footnotesize]
{"composition": [ 
 {"name":"get_parking-e-available", 
    "params":["p1","b1"]},
 {"name":"post_book-parking-e", "params": 
    ["p1", "r1", "b1", "m1"]},
 {"name":"book-tirepressurecheck", 
    "params":["p1", "m1", "r1"]},
 {"name":"get_parking-navigation-parkingid", 
    "params":["p1"]}], 
"environment": [ 
 {"value":"", "type":"parkingid", "name":"p1"}, 
 {"value":"", "type":"operatorid", "name":"b1"},
 {"value":"", "type":"reservationnr", "name":"r1"},
 {"value":"", "type":"maxparkingtime", "name":"m1"},
 {"value":"", "type":"bookedservice", "name":"g1"}]}
\end{lstlisting}
\caption{Formalized Composition Result.}
\label{fig:listing2}
\end{figure}

\section{Conclusion and Future Work}\label{sec:conclusion}

We presented a solution architecture for an Emergent Software-Service Platform and  
a smart mobility use case for which a prototypical implementation of the platform exists.
First experiments with the existing prototype showed that it is able to automatically elicit the user requirements from an explicit request and is able to automatically compose and execute a software service that meets the recognized user requirements.
Hence, these results show that it is viable to interpret the problem of autonomous self-adaptive software-service composition as a classical planning problem.

Nevertheless, the presented evaluation use case and the prototypical implementation of course still have some limitations.
One limitation is that the considered use case is rather small, and only one possible form of interaction between a user and the software platform (i.e., via the Request Configurator) was evaluated.
In future research, it would be interesting to extend this to other, possibly even more complex, forms of interaction like recognizing user requirements directly from natural language.
Another limitation is that it is not possible to roll back the execution of the composed software service when the platform detects at execution time that the calculated software composition does not work.
This might happen as it cannot be checked during the composition process whether the software services that are composed into a new  service will be available at execution time.
Also interesting to consider for future work is to implement and evaluate a more sophisticated feedback loop between the software service platform and the user(s).
This would potentially not only increase the acceptance of the users to use such a platform but also enables the possibility for the platform to learn from more detailed user feedback to improve future software service compositions.


\section*{Acknowledgment}
This work was funded by the German Federal Ministry of Education and Research (Research Grant: 01IS18079, Project: BIoTope) and the German Federal Ministry for Economic Affairs and Climate Action (BMWK) (Research Grant: 01ME\-21002, Project: HitchHikeBox).

\bibliographystyle{IEEEtran}
\bibliography{biotope_toolpaper}

\end{document}